**Extended electrochemical monitoring of biomolecular binding using commercially available, reusable electrodes in microliter volumes.**

Jeremy Mendez[*1], Yae Eun Kim[*1], Nafisah Chowdhury[*1], Alexios Tziranis[*1], Phuong Le[*1] Angela Tran[*1], Rocio Moron[*1], Julia Rogers[*1], Aohona Chowdhury[*1], Elijah Wall[*1], Netzahualcóyotl Arroyo-Currás[2], Philip Lukeman[X1]

*=Undergraduate co-authors, X=corresponding author lukemanp@stjohns.edu
1= Department of Chemistry, St. John's University, NY
2= Department of Pharmacology and Molecular Sciences, Johns Hopkins University School of Medicine, MD

E-DNA sensors (Fig 1A) are the oligonucleotide-detecting precursors to Electrochemical Biosensors (EBs). These EBs utilize oligonucleotides (including aptamers) as the molecular recognition component of the sensor to detect DNA, small molecule and protein analytes. To construct these sensors, the oligonucleotide is synthesized with redox reporters (a functional group that can be oxidized or reduced at an appropriate potential) at one end of the strand; the most commonly used of these reporters is Methylene Blue (MB). The distal end of the oligonucleotide is functionalized with an alkylthiol; this thiol enables attachment of the oligo to a gold electrode surface, forming a mixed monolayer with added small-molecule alkylthiols (such as 6-mercaptohexanol, MCH), which serve to protect the electrode surface from nonspecific binding and oxygen. These electrodes, when placed in an 3-electrode electrochemical cell in contact with a solution of analyte, act as sensors. The do so by transducing binding events that result in conformational change - and these changes' resultant effect on redox reporter electron transfer kinetics - into an electrochemical signal. These sensors have been utilized in everything from buffer solutions to living animals allowing real-time monitoring of analyte concentration in these challenging environments, and are on a path to commercialization.[1]

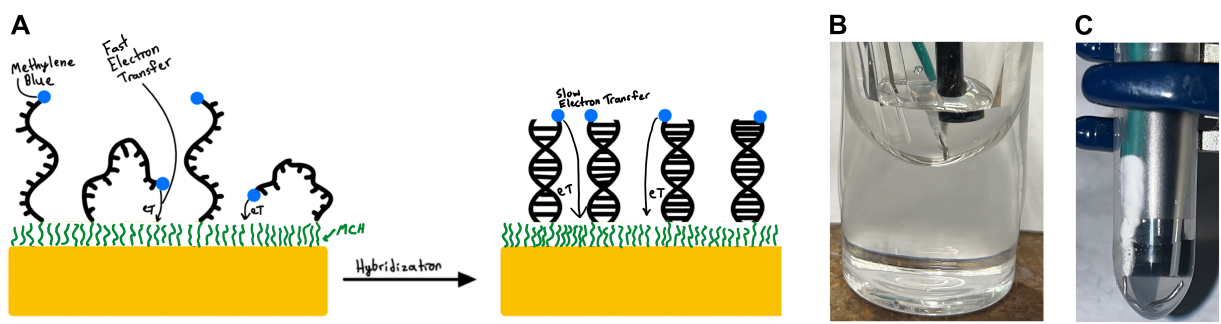

**Fig 1**. A) An E-DNA sensor used to detect oligonucleotide hybridization. B) "Conventional" E-DNA sensor setup with a gold working electrode, Pt wire counter electrode, a macroscale liquid Ag/AgCl reference electrode and shot glass as cell. C) Our adaptation using a leakless microscale reference electrode and a "low-bind" Eppendorf-brand microcentrifuge tube as cell.

Historically, sensor development and optimization has originated in academic labs with a few standard experimental configurations for the electrochemical cell and data analysis. We report here on adaptations of these that are friendly to novice scientists such as those in undergraduate laboratories.

The standard electrochemical cell that is most utilized within these academic labs, that allows reuse of expensive gold electrodes is as follows: The working electrode is a 2 mm diameter Gold Rod in a ~6 mm diameter Kel-F polymer sheath, the counter-electrode is a Pt Wire, and the reference electrode a liquid Ag/AgCl reference electrode in a ~6 mm diameter glass cell with porous frit, with a "shot glass" serving as the electrochemical cell holder and electrolyte[2] (Fig 1B). Voltametric measurements are conducted with a research-grade potentiostat such as those by Metrohm, Gamry or CH instruments. We describe alternate approaches to this platform - and their disadvantages - from the literature and other academic labs in Appendix 1.

While serving the field well, the standard as configured has limitations;
i) the reference electrode when immersed in electrolyte can leach both silver[3] and chloride[4] ions over time, which can profoundly affect electrochemical measurements.
ii) the diameters of the reference and working electrode preclude the use of low volumes of analyte solution; especially challenging for expensive or rare analytes.
iii) glass vessels bind to biomolecules – especially DNA, causing sample loss and measurement errors.
iv) research-grade potentiostats are expensive; their software is aimed at expert users and the GUIs powering them, and software that can generate statistical analyses are quite novice-unfriendly, and require python skills to use and compile.

We were motivated to overcome these difficulties in a project that our undergraduate research group developed, *which required use of small amounts of DNA analyte at low concentrations.* This work utilized the self-assembled DNA 5-pointed star (5PS) developed by Mao[5] as a model system that could be used to investigate polyvalent binding on electrode surfaces; other groups have come to the same conclusion about this system's utility for investigation of polyvalent surface binding[6]. Using Square Wave Voltammetry (SWV), we investigated the effect of polyvalent binding of 5PS on electron transfer rate of an E-DNA sensor using Lovrić's formalism[7]; we plotted frequency-normalized peak intensity *vs.* log frequency to determine this rate. The hypothesis behind this work was that we could differentiate monovalent vs. polyvalent binding if the distance between the binding sites on the 5PS and the binding sites on the surface matched size, resulting in a reduction in flexibility if and only if upon polyvalent binding occurred. We have observed this effect before – polyvalent binding with matched size surface-bound DNA origami receptors reduced electron transfer rates[8]. Experimental limitations turned out not to allow this analysis for this system (this analysis will be subject of a future publication), but in our efforts we overcame the reference electrode, sample size, cost and software limitations described above, to generate reliable electrochemical binding data, in an undergraduate-friendly manner.

In brief our protocol is as follows. We change the electrochemical cell and reference electrode, potentiostat and analysis workflow from the standard setup. The electrochemical cell is a commercially available microcentrifuge tube (DNA or protein Lo-Bind, Eppendorf); we have worked with volumes of electrolyte/analyte as low as 150 µL. These microcentrifuge tubes have been shown to have low (<5%) sample loss of low concentration solutions of Protein[9] and DNA[10] solutions without the need for surfactants or other additives. The reference electrode is a leak-free polymer electrode (LF1-45, innovative instruments); its sheath is made of PEEK polymer (polyether ethyl ketone) has a diameter of 1 mm. (Fig 1C)

The working electrodes are as described in the literature – gold surfaces displaying DNA oligos with the surface protected by a MCH monolayer; reusable between experiments by mechanical and electrochemical cleaning. Usually, multiple measurements in one experiment are conducted on these electrodes by washing with water/buffer. The polyvalent 5-pointed stars we bound to the surface are quite strongly bound; conventional washing with water does not entirely remove them; we did not want to use heat or more aggressive pH change for fear of damaging the monolayer or surface density of oligos. By adapting a protocol described for protein biosensors[11] we resolved this issue; washing with 8M aqueous urea completely removes bound DNA. When the washing solution is doped with a small quantity of MCH, this backfills the monolayer which is damaged by both washing and electrode interrogation over time. We have performed repeated experiments on individual electrodes – measure voltammograms, perform kinetic analysis of binding over hours and wash electrode - repeated up to 5 times on one electrode; with the number of scans and our voltage regime only being slightly damaging to the monolayer[12], usually the length of the working day is the limiting factor for use of one preparation of these electrodes. We note that electrode *positioning* as described in Appendix 2 is crucial - the geometries of working, reference and counter electrodes when analyses are run in small volumes has been shown to be important with other E-AB systems[13].

We utilize potentiostats from PalmSens (EmStat and PalmSens4) these are between 3 to 10 times cheaper than potentiostats from other suppliers above when purchased new. Older EmStats, which are sensitive enough for SWV with these electrodes, from 1 - 400 Hz in our hands, can be obtained on the used market for less than $200. Critically, unlike some older used potentiostats, the analysis software, which works across every generation of potentiostat is actively supported by manufacturer to this day, has a robust and novice-friendly GUI which works on modern PCs and allows scripting of multiple experiments with point-and-click.

We created software in Microsoft Excel that requires no knowledge of coding that allows the raw current/voltage output of Palmsens software (Fig 2A) to enable generation of "Lovrić plots" (Fig 2B). Background-subtraction based on a rolling average of the voltammogram baseline is performed before the plot is generated. Multiple series of these plots can be overlain on the same graph (Fig 2C) - inter experiment and inter-electrode variability can be displayed (Fig 2D). We found that with this system and analytical tool, we could interrogate electrodes as reliably and reproducibly as comparable systems described in the literature[14]; while the majority of our data was obtained with E-DNA sensors, we found that aptamers to aminoglycosides worked equally as well.

All instructions - in the form of a procedural guide - on how to prepare electrodes in this manner, as well as links to the Excel software are in Appendix 2.

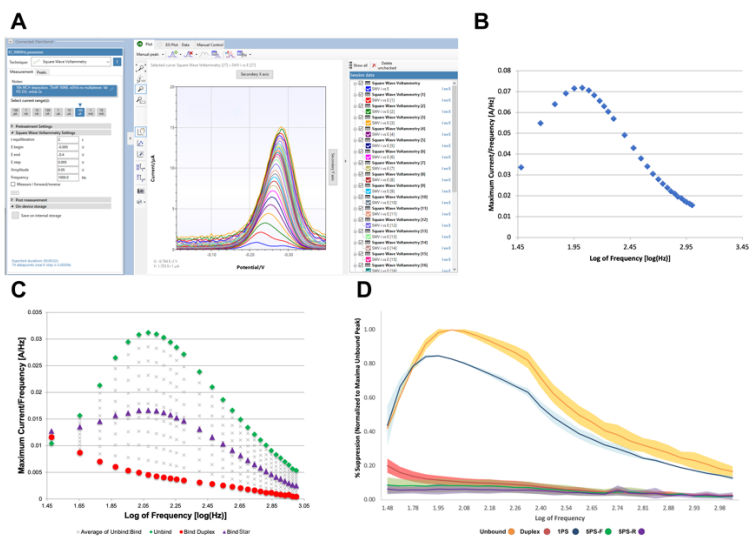

**Figure 2.** A. PalmSens potentiostat software that shows its user-friendly GUI with an overlapping series of Square Wave Voltammograms . B. Output of a single "Lovrić plot" from our custom built excel sheet. C. Extrapolating weighted means (grey X's) of two Lovrić plots (red and green) allows a visual comparison of a third plot (purple) to determine similarities. D. Statistical analyses of multiple Lovrić plots from 3 biological replicates on multiple samples showing standard deviation as shaded lines.

**Future work**

To extend this work to even smaller volumes, the limiting factor for this current setup is the commercially available gold-rod electrode. The likely minimum size of the sheath is 0.4 mm, meaning that depending on gold rod area, a 0.9 mm electrode would be feasible. This would allow 3 or 4 electrodes in a 2 ml microcentrifuge tube, or potentially one electrode in a narrower-bottomed conical tube, which we extrapolate to enabling measurements in 70 μL. We have experimented with 3d printed rod electrodes for this task - work to this end is ongoing. While we have tested this system with square-wave voltammetry between 5-800 Hz, and conventional cyclic voltammetry, other electrochemical interrogation techniques[15],s[16] may be affected by small volumes and should be examined further.

We further expect that this work can be combined with other novice-friendly processes such as 3d printing of electrodes[17] and microfluidic electrochemical biosensors[18] , thus allowing for the wider use of electrochemical biosensors in undergraduate laboratories.

**Acknowledgements**

We thank Martin Kurnik and Kaylyn K. Leung for helpful discussions. P.S.L. is grateful for support from Army Research Office (awards W911NF-19-1-0326 and W911NF-23-1-0283).

**Appendix 1.** Alternate approaches to our techniques

There are other electrochemical cell/electrode/potentiostat combinations which address some but not all of the issues we solved with our approach.

**Potentiostats and software**: While *many* open-source homebuilt potentiostats have been described,[19] they require some microelectronics experience to construct and both the software and devices are unsupported beyond a volunteer community. We note that powerful software [20] has been written that takes the output of research-grade potentiostats and can perform the many analyses including the ones described in this, including in real-time; however, the workflow for the novice is daunting, and using it requires some knowledge of Python to install and execute.

**Working electrodes**: As biosensor electrodes have been utilized in living animals[13], the field has moved in the direction of using the same gold-wire setup for working electrode screening as *in vivo*. While affordable, they are not reusable and require soldering and construction skills.

The use of commercially screenprinted[21] electrodes solves the volume problem; they can be quite affordable, but they are not reusable; for gold working electrodes as described here, the quality of the gold on these surfaces is also dependent on the print run. In addition most of these electrodes are open to the air and rely on capillary forces to hold an electrolyte/analyte drop in place which is prone to evaporation. Constructing or buying sealed/evaporation-proof systems increases complexity and cost.

Lithographically printed electrochemical cells enable single or parallelized investigation of sensors in low volumes[22]; similarly ultramicroelectrodes capable of microliter measurements[23] have been constructed. These electrodes are non-reusable, delicate, expensive, and require fabrication skills beyond that of an undergraduate lab. While progress in 3d-printing has simplified micro-electrode construction that are capable of making multiple measurements in parallel[14], these approaches still require fabrication efforts.

**Reference Electrodes**: A much smaller leakless reference electrode[24] has been fabricated; this is quite delicate, require manufacturing skills and are is not reusable.

**Appendix 2.** Procedural guide to electrode preparation and analysis

Excel sheet, scripts and methods available here:
https://drive.google.com/drive/folders/1bva1yzCUmgRo95G72zPto3qcXQQKtqv7?usp=drive_link
Chemicals were reagent grade from fisher unless stated otherwise.

**Solution Recipes:**

| | |
|---|---|
| 1mM MCH | 1.3uL of MCH (Sigma Aldrich, 99%+) in 10mL of 1X PBS |
| 5mM TCEP.HCl/PBS | 0.014g of TCEP.HCl in 10mL of 1X PBS |
| Wash Solution | [8M Urea (4.8g of Urea completely dissolved in DD water) + 500uL of 1mM MCH] in 10mL of DD water |
| 10X HEPES MgCl | 50mL 1M HEPES ph 7.4 + 25mL 1M $MgCl_2$ (50.83g in 250mL of DD) + 175mL of DD Water |
| 1X PBS | 1 Tablet + 200mL of DD Water |
| 0.5M $H_2SO_4$ | 13.9mL $H_2SO_4$ in 500mL DD Water |
| 0.5M NaOH | 10g NaOH in 500mL DD Water |
| 1:1 EtOH:H2O | 200mL of Ethanol + 200mL of DD water |
| 1X PPC | 3.35g $NaClO_4$ + 0.0408g KCl + 0.287g dibasic $Na_2HPO_4$ + 0.0453g monobasic $NaH_2PO_4$ in 200mL of DD water |
| 5mM TCEP.HCl/PPC | 0.014g of TCEP.HCl in 10mL of 1X PPC |

**DO's and DON'T's for Solutions:**

| DO | DON'T |
|---|---|
| <ul><li>Label all Solution Containers with actual recipe and DATE</li><li>**Every start of the week**, make *NEW*:<ul><li>MCH</li><li>TCEP.HCl</li><li>WASH</li></ul></li><li>Keep 10X HEPES.Mg in aluminum foil and make sure to take a pH reading (from bottle or with probe)</li><li>Keep MCH and TCEP.HCl in foil and in the fridge</li><li>Stir NaOH before use</li><li>Make **NEW 10X HEPES EVERY 3 WEEKS**</li><li>Use new FILTER every time</li></ul> | <ul><li>***DO NOT use +6 month old MCH***</li><li>DO NOT use +3 week old 10X HEPES</li><li>DO NOT keep MCH, TCEP.HCl out of the fridge</li><li>DO NOT make $H_2SO_4$ without PPE<ul><li>***WATER BEFORE ACID***</li></ul></li><li>DO NOT use plastic pipette for conc. $H_2SO_4$ (glass only!!)</li><li>DO NOT use first couple drops of filtering solution (dispose first couple drops cuz of glycerine on filter)</li><li>DO NOT empty waste into DD water beaker (only water plz)</li><li>DO NOT spill MCH (smell sticks/lingers if not cleaned properly)</li><li>DO NOT use MCH over 6 months old (will give bad results- can be seen by washes not performing the same as the initial surface readings)</li></ul> |

**Working Electrode Storage and Maintenance:**
- Rinse finished reference electrode with DD water and dry with kimwipe
- Store in cool dark place
- Make sure working electrode:
  - Have <u>no cracks/scratches</u> in the <u>gold surface</u>
  - Have <u>no DEEP cracks</u> in the <u>polymer casing</u>

*Advice: If the electrode gives strange results, figure out whether sandpapering is needed

**Reference and Counter Electrode Storage and Maintenance:**
*'Fat' Reference Electrode (RE)*
- Make sure:
  - KCl solution is full the electrode; should be below the Teflon cap
  - Rod inside is mostly **brown**; NOT shiny
  - There is solution inside the glass rod; must be **completely filled**
  - **Frit is clean**; doesn't look yellow/gunky

- <u>How to Refill Glass</u>:
  - Use 3mL Syringe + Hypodermic Needle
  - Gently twist the Teflon cap off the glass tube (DO NOT PULL)
  - Use syringe to pull out the old KCl fluid, make sure no fluid is in Teflon cap
  - Once empty, use a syringe to refill both the cap and the glass tubing; must be completely full for both cap and tube
  - Tap the tubing, make sure there are no air bubbles in the tube
    - SLOWLY fill the solution with the needle close to the bottom of the tube
    - DO NOT poke the frit with the needle
  - SLOWLY insert the glass tubing into the Teflon Cap all the way in
    - You will see solution coming out through the frit; that is normal
  - Let it sit until no more solution comes out through the top
  - Put it in ~4-5mL of KCl
- <u>How to Clean Frit</u>:
  - Soak frit in 0.1M HCl for 30 mins
  - Sonicate the frit for 10 s in 0.1M HCl
  - Immerse in DD water for 10 mins

*'Skinny' – 1LF - Reference Electrode*
- Storage in small vial so that **only the tip is submerged** in solution
  - Solution = 0.05M $H_2SO_4$ with few drops of KCl
- If the potential shifts from 0.26V for the SWV readings, proceed with cleaning:
  - Soak in 0.5M $H_2SO_4$ overnight, if that doesn't work…
  - Soak in acetone for 2hrs, if that doesn't work…
  - Polish with extremely lens paper or weighing paper for 30 seconds
  - Soak in ethanol and $H_2SO_4$ mix for 30 seconds

- Place the cleaned electrode in a new solution for storage after rinsing with DD water

*Counter Electrode*

Are there issues not addressed by the above? Look at counter
- Take off Teflon tape
- Rewrap exposed copper wires around the gold rod
- Wrap with Teflon tape
- Rinse with DD water before storing in 50mL tubes

**Polish and Clean electrodes:**

*Prepping for Polishing*
  i. Clean shot glasses with DD water (+soap if it looks crusty)
 ii. Make sure 1:1 EtOH:H2O is not >3 Weeks old
iii. Rinse ALL shot glasses with **solution** before use
iv. Make sure petri dish pads aren't demolished; replace as needed

*Polishing*

  i. Polish working electrode(s) (WE) in a figure 8 pattern for ~2 min (timer!) on a cloth pad with **MetaDi-diamond polish**.
 ii. Rinse with DD water from a squirt bottle.
iii. Sonicate for 30s in fresh 1:1 EtOH/Water.
iv. Rinse with DD water from a squirt bottle.
 v. Polish WE(s) in a figure 8 pattern for ~2 min (timer!) on a cloth pad with 0.05 μm **alumina/water slurry.** (*Add more DD water if the pad is too dry or alumina if needed*)
vi. Rinse with DD water from a squirt bottle.
vii. Sonicate for 30s in fresh 1:1 EtOH/Water
viii. Rinse with DD water from a squirt bottle.
   #Polished WE(s) can be stored in DD water while waiting for electrochemical cleaning.

*Prepping Electrochemical Cell*
  i. Use clean shot glasses
 ii. Make sure to clip all the electrodes: working (red); counter (black); reference (blue)
iii. Make sure PSTrace is ON and CONNECTED (check PSTrace Software Interface)
iv. Rinse and fill will solution; Small tall small shot glasses fill more than halfway

*Cleaning*
i. Clean in 0.5 M NaOH (Fresh and stirred)
Method>Load>Follow * to select **NaOH clean 1**
Rinse with DD water and then 0.5 M H2SO4

  ii.    Clean in 0.5 M H2SO4 (Fresh)
Method>Load>Follow * to select **H2SO4 clean 1**

#Polished WE(s) can be stored in DD water while waiting for checks.
DO NOT leave in H2SO4 unattended for too long (no more than 5 minutes).

*Checking Surface*
i. Check in FRESH solution 0.5 M H2SO4
[Desktop/NEW Lukemanlab pstrace/Cleaning Scripts/]*
Method>Load>Follow * to select **H2SO4 CHECK 1**
    ii.    Rinse with DD water before putting into a shot glass contained fresh 0.05 M H2SO4.
    ii.    Check in 0.05 M H2SO4 for area determination
[Desktop/NEW Lukemanlab pstrace/Cleaning Scripts/]*
Method>Load>Follow * to select **0.05M H2SO4 CHECK**
    iv.    Save this file!

Error in reading, try: (1) fresh solution, (2) different RE, (3) different CE. If not, NOXXXX

**Functionalization and Deposition:**

*If cleaning voltammogram acceptable...*
i. Rinse with DD water
ii. Shake to dry
iii. Place immediately into Surface Deposition Solution
iv. ***Agitate*** WE in solution
v. Leave WE in solution **for 1 hour**
vi. Move WE immediately into **MCH without rinsing** with DD Water
vii. Leave overnight
viii. Take out of MCH to keep deposition times the same
ix. Rinse with ***FILTERED BUFFER***
x. Place ***in 1X FILTERED BUFFER*** and equilibrate for 10 mins

**Measurements**

*Density*
CV
 i. Connect WE to a working cell
ii. Run CV
  Method>Load>Follow * to select **Determination of Packing Density**
  #Run CV after equilibrating in buffer in a cleaned Shot Glass

*Density Calculation*
  Calculate Area of Gold
  i. Load voltammogram (0.05M H2SO4 check)
 ii. Select LAST voltammogram (voltammogram 3)
iii. Select integration*
iv. Draw line from 0.6V to 0.0V
 v. Take down integration value in **Density Formula.xlsx** under **H2SO4** column (change formula to divide by 2 electrons for ferrocene measurements)

  Calculate Quantity of E-DNA on surface

  i. Load voltammogram (Density check)
 ii. Select LAST voltammogram (voltammogram 3)
iii. Select integration*
iv. Draw line from straight line before curve to straight line after curve
 v. Take down integration value in **Density Formula.xlsx** under **CV** column

* 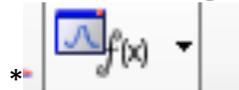

**Electrode Positioning for SWV in Eppendorf**
*Heuristics*
 i. Place WE in 2mL Eppendorf tube (DNA Lobind)
 ii. Attach alligator wire CE (counter) onto the black wire so that the clips are perpendicular to each other
 iii. Place CE into the tube; take out the WE just slightly so it can fit
 iv. Make sure the platinum wire tip is lower than the WE surface
 v. Clip the red wire to the WE
 vi. Slip the blue wire to be in between the black and red wire before clipping to the THIN RE
 vii. Gently slip the clipped THIN RE into the tube
 viii. Make sure
  a.   the tip of the RE is submerged in solution
  b.   the RE is not too close to the CE
  c.   the tip of the RE is higher than the WE surface
 **REFER TO THIS DIAGRAM:**

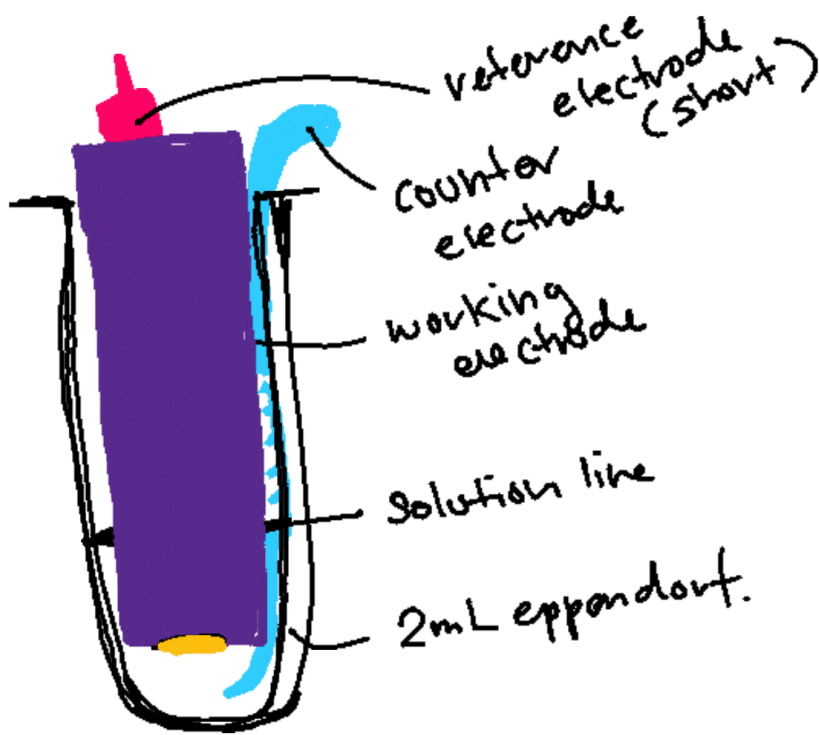

*SWV*

i. After obtaining a good overlapping baseline, run SWV script.
   [Desktop/NEW Lukemanlab pstrace/Electrode Scripts/]*
   Script>Open script window…>File>Load script>Follow * to select **E#_30Hz_to_1000Hz**
   > Measurement (Frequency 30.0000Hz) >Edit>**Change Note Content as Needed**>Save&Close
   > Measurement (Frequency 1000.00Hz) >Edit>**Change Note Content as Needed**>Save&Close

**SWV Note Format:**
[Time] MCH, [Surface Content], E[name], [no/yes] multiplexer, [Thin/Thicc] RE [name],
[Additional Notes ie. Bind, initial, wash, etc. etc.]

- ii. Run!
- ii. Move all files in **E1 Temp Folder** to newly named folder
- ii. <Export as CSV file> and save in **SWV Files for Export** Folder

**\*\*WATCH OUT!!**

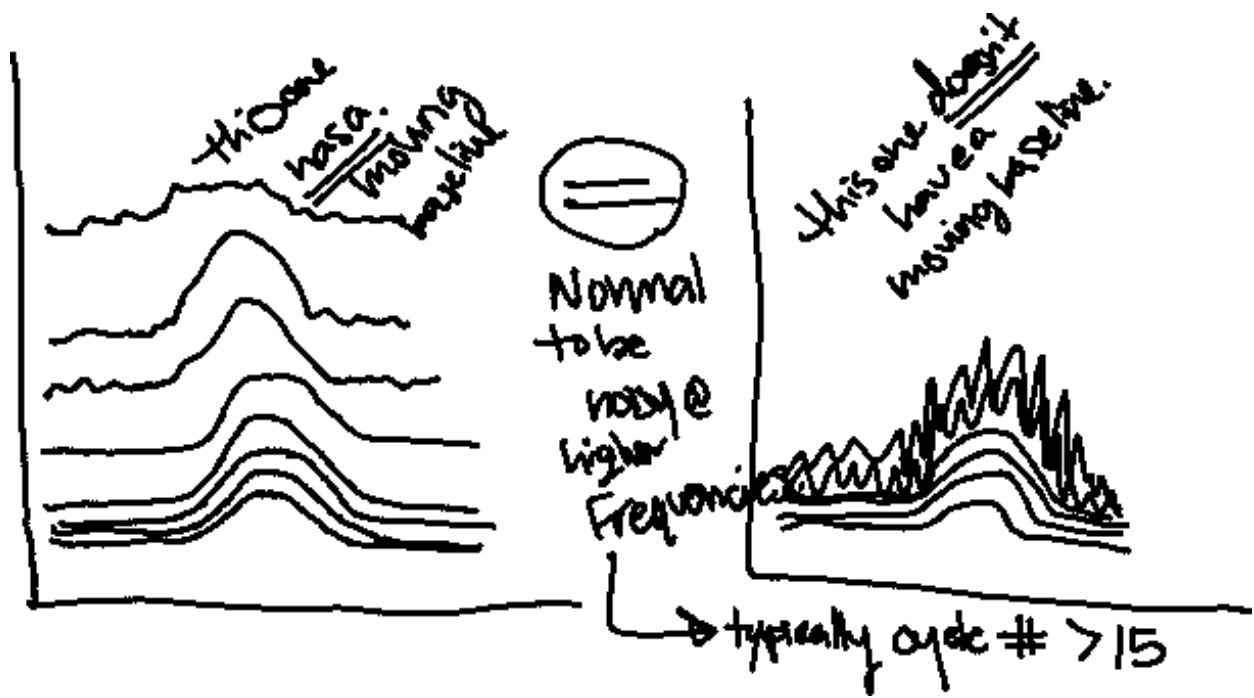

**HOWEVER!!**
- Walking Peaks    **BAD**
- Extremely Slanted Baselines    **BAD**
- Dropping Baselines    **BAD**
- SUPER Noisy Peaks    **BAD**

**Data Analysis (Excel Sheets for SWV)**:
Files→ Student Data- Summer→ Coding Files (choose version 5.8 or 5.9 of PS Trace)
1. Rename sheet with date and save
2. Import data from flash drive
a. Select all (Command + A), Copy
b. Paste into Ref tab
c. Copy lane from Graphing tab
d. Paste into Comparisons tab

***Save the sheet you are working on often as excel is prone to crashes***